\documentstyle[prb,aps,epsf,floats]{revtex}
\begin{document}
\def\lesssim{\mskip\thickmuskip\raise+1pt\hbox{$<$}\mkern-14.mu\lower+3.5pt\hbox{$\sim$}\mskip\thickmuskip}
\renewcommand{\textfraction}{0.0}
\renewcommand{\floatpagefraction}{.7}
\setcounter{topnumber}{5}
\renewcommand{\topfraction}{1.0}
\setcounter{bottomnumber}{5}
\renewcommand{\bottomfraction}{1.0}
\setcounter{totalnumber}{5}
\setcounter{dbltopnumber}{2}
\renewcommand{\dbltopfraction}{0.9}
\renewcommand{\dblfloatpagefraction}{.7}

\draft

\twocolumn[\hsize\textwidth\columnwidth\hsize\csname@twocolumnfalse%
\endcsname

\title{Griffiths Singularities in the Disordered Phase of a  \\
Quantum Ising Spin Glass}

\author{H. Rieger}

\address{HLRZ, Forschungszentrum J\"ulich, 52425 J\"ulich, Germany\\
Institut f\"ur Theoretische Physik, Universit\"at zu K\"oln,
50937 K\"oln, Germany}


\author{A. P. Young}
\address{Department of Physics, University of California, Santa Cruz,
CA 95064}
\date{\today}

\maketitle

\begin{abstract}
We study a model for a quantum Ising spin glass in two space dimensions
by Monte Carlo simulations. In the disordered phase at $T=0$, we find power law
distributions of the local susceptibility and local non-linear
susceptibility, which are characterized by a smoothly varying dynamical
exponent $z$. Over a range of the disordered phase near the 
quantum transition,
the local non-linear susceptibility diverges. The local susceptibility
does not diverge in the disordered phase but does diverge at the
critical point. Approaching the critical point from the disordered
phase, the limiting value of $z$ seems to equal its value precisely at
criticality, even though the physics of these two cases seems rather
different
\end{abstract}
\vskip 0.5 cm
]

\section{introduction}
A feature of disordered systems which has no counterpart in pure
systems, is that rare regions, which are more strongly correlated than
the average, can play a significant role. For classical magnetic
systems, Griffiths\cite{griffiths} showed that such regions lead to a
free energy which is a non-analytic function of the magnetic field at
temperatures below the transition temperature of the pure system.
However, for the static properties of a classical system, the
Griffiths singularities are very weak; just essential
singularities\cite{essen}.  By contrast, Griffiths singularities are
much more spectacular for quantum phase transitions at $T=0$,
especially for systems with a broken {\em discrete} symmetry.  One
model where these effects can be worked out in great detail, and where
Griffiths singularities dominate not only the disordered phase but
also the critical region, is the one-dimensional random
transverse-field Ising model\cite{mw,sm,dsf,yr}. In that model, one
finds very broad distributions of various quantities including the
local susceptibility and the energy gap, and a dynamical exponent,
$z$, which is infinite at the critical point. In the disordered phase,
the distribution of local susceptibilities is found to be a power law,
in which the power can be related\cite{yr} to a continuously varying
dynamical exponent\cite{dsf}, which diverges at criticality. The {\em
average} susceptibility diverges when $z > 1$, i.e. over a finite
region of the disordered phase in the vicinity of the critical point,
a result first found many years ago by McCoy and Wu\cite{mw}.

Many of these surprising results, such as the power law distribution
of susceptibilities in the disordered phase, are expected to hold more
generally\cite{th}. However, it is not clear whether the average
susceptibility will diverge in the disordered phase for dimension,
$d$, greater than 1, or whether this is a special feature of
$d=1$. Here we investigate Griffiths singularities for a {\em
two\/}-dimensional quantum Ising spin glass system by Monte Carlo
simulations.  Additional motivation for our study comes from
experimental work\cite{expt} on a quantum spin glass system, which did
not find the expected strong divergence in the non-linear
susceptibility at the quantum phase transition. By contrast,
subsequent numerical simulations\cite{ry,gbh} did find a rather
strong divergence, comparable with that at the classical spin glass
transition. Hence it seems worth investigating whether this
discrepancy might be due to Griffiths singularities causing the
non-linear susceptibility to diverge even in the disordered phase,
thus making the location of the transition difficult in the
experiments. Somewhat less detailed
results on the two-dimensional spin glass have also
been reported in parallel work by Guo et al.~\cite{gbh2}, who,
additionally, performed calculations in three dimensions. 

\section{the model}

The two-dimensional quantum Ising spin glass in a transverse
field~\cite{Chakrabarti} is
defined via the following quantum mechanical Hamiltonian
\begin{equation}
\tilde{\cal H}_{QM} = - \sum_{\langle i,j\rangle} 
\tilde{J}_{ij} \sigma_i^z\sigma_j^z - \Gamma\sum_i \sigma_i^x\;,
\label{hamqm}
\end{equation}
where the $\{\sigma^\alpha_i\}$ are Pauli spin matrices,
the $\tilde{J}_{ij}$ are
quenched random interaction strengths and $\Gamma$ is an external
transverse field. A system described by this Hamiltonian undergoes a
quantum phase transition at zero temperature, $T=0$, from a
paramagnetic (or spin liquid) phase to a spin glass phase for some
critical field strength $\Gamma_c$ \cite{ry}. As is described
elsewhere \cite{ry,gbh} this model can be mapped onto an effective
classical Hamiltonian in two space plus one imaginary time dimensions,
with disorder that is perfectly correlated along the imaginary time
axis.  This classical Hamiltonian is
\begin{equation}
{\cal H} = -\sum_\tau \sum_{\langle i,j\rangle} J_{ij} S_i(\tau)
S_j(\tau) -\sum_{\tau, i} S_i(\tau) S_i(\tau + 1) ,
\label{ham}
\end{equation}
where the Ising spins, $S_i(\tau)$, take values $\pm 1$, $i$ and $j$
refer to the sites on an $L \times L$ spatial lattice, while $\tau$
denotes a time slice, $\tau = 1, 2, \ldots , L_\tau$. The number of time
slices, $L_\tau$, is proportional to the
inverse of the temperature, $T$, of the original
quantum system in Eq.~(\ref{hamqm}). Periodic
boundary conditions are applied in all directions.  The model is
simulated at an effective classical ``temperature'' $T^{\rm cl}$,
which controls the amount of order in the spins. Of course, $T^{\rm cl}$,
is not the real temperature $T$, which is zero at the
transition, but is rather a measure of the strength of the {\em
quantum} fluctuations. Increasing $T^{cl}$ therefore corresponds to
increasing the transverse field $\Gamma$, in the quantum Hamiltonian,
Eq.~(\ref{hamqm}). 
The nearest neighbor interactions $J_{ij}$ are independent of $\tau$,
because they are quenched random variables, and are chosen
independently from a Gaussian distribution with zero mean and standard
deviation unity, i.e.
\begin{eqnarray}
[J_{ij}]_{\rm av} & = & 0, \nonumber \\
\ [J_{ij}^2]_{\rm av} & = & 1 \ ,
\end{eqnarray}
where $[ \cdots ]_{\rm av}$ denotes an average over the disorder.
Statistical mechanics averages for a given sample will be denoted by
angular brackets, i.e. $\langle \cdots \rangle$. 
The interactions between time slices are ferromagnetic and taken to be
non-random with strength unity.

\begin{figure}[hbt]
\epsfxsize=\columnwidth\epsfbox{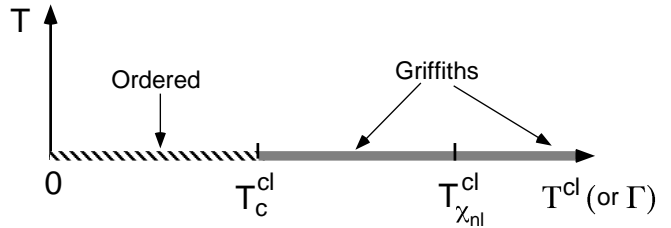}
\caption{
The phase diagram of the two-dimensional quantum Ising spin glass.  The
horizontal axis can be thought of as $T^{\rm cl}$ if one is using the
effective classical Hamiltonian in Eq.~(\protect\ref{ham}) or $\Gamma$
if one is using the original quantum Hamiltonian in
Eq.~(\protect\ref{hamqm}).  There is a critical point at $T^{\rm cl} =
T_c^{\rm cl}$.  For $T^{cl} < T_c^{\rm cl}$ there is a spin glass
ordered phase. For $T^{cl} > T_c^{\rm cl}$ there is no spin glass order
but there are Griffiths singularities. In the region $T_c^{\rm cl} <
T^{\rm cl} < T_{\chi_{\rm nl}}^{\rm cl}$ the Griffiths singularities
are sufficiently strong that the average non-linear susceptibility
diverges as the (real) temperature tends to zero. For $T^{\rm cl} >
T_{\chi_{\rm nl}}^{\rm cl}$ the average non-linear susceptibility stays
finite in the zero temperature limit.
}
\label{figm1}
\end{figure}

The phase diagram of the model is sketched in Fig.~\ref{figm1}.
Because we are dealing with a two-dimensional lattice, there is no
finite-temperature spin glass transition\cite{review}, and the region
with spin glass order therefore lies entirely along the $T=0$ axis in
the region $0 \le T^{\rm cl} < T_c^{\rm cl}$, where
$T_c^{\rm cl}$ denotes the critical point.
In earlier work\cite{ry} we found that
\begin{equation}
\label{tc}
T^{\rm cl}_c = 3.275 \pm 0.025 \ .
\end{equation}

\section{theory}
Griffiths singularities arise from localized regions which are more
correlated than the average. They do not have large spatial
correlations, but give rise to singularities because of large correlations
in imaginary time. To focus on these time correlations, it is
simplest to study quantities that are completely {\em local} in space, i.e.
are just on a single site. We shall be particularly interested in the local
susceptibility,
\begin{equation}
\chi^{\rm (loc)} = {\partial \over \partial h_i} \langle \sigma_i^z
\rangle ,
\end{equation}
where $h_i$ is a local field on site $i$. In the imaginary time
formalism this can be evaluated from
\begin{equation}
\chi^{\rm (loc)} = 
\sum_{\tau = 1}^{L_\tau} \langle S_i(0) S_i(\tau) \rangle \ .
\label{defchi}
\end{equation}

Since the divergent response function at a conventional spin glass
transition is the non-linear susceptibility\cite{review},
it is also interesting to
study the local non-linear susceptibility, given by
\begin{equation}
\chi^{\rm (loc)}_{\rm nl} = {\partial^3 \over \partial h^3_i} \langle \sigma_i^z
\rangle ,
\label{chinlqm}
\end{equation}
which can be determined in the simulations from
\begin{equation}
\chi^{\rm (loc)}_{\rm nl} = -{1 \over 6 L_\tau} \left( \langle m_i^4 \rangle
- 3 \langle m_i^2 \rangle^2 \right)  \ ,
\label{defchinl}
\end{equation}
where
\begin{equation}
m_i = \sum_{\tau = 1}^{L_\tau} S_i(\tau) \ .
\end{equation}

We consider distributions of $\chi^{\rm (loc)}$ and $\chi^{\rm (loc)}_{\rm nl}$
obtained both by measuring at different sites in a
given sample, and by taking many samples with different realizations
of the disorder.

\subsection{Disordered Phase}
\begin{figure}[hbt]
\epsfxsize=\columnwidth\epsfbox{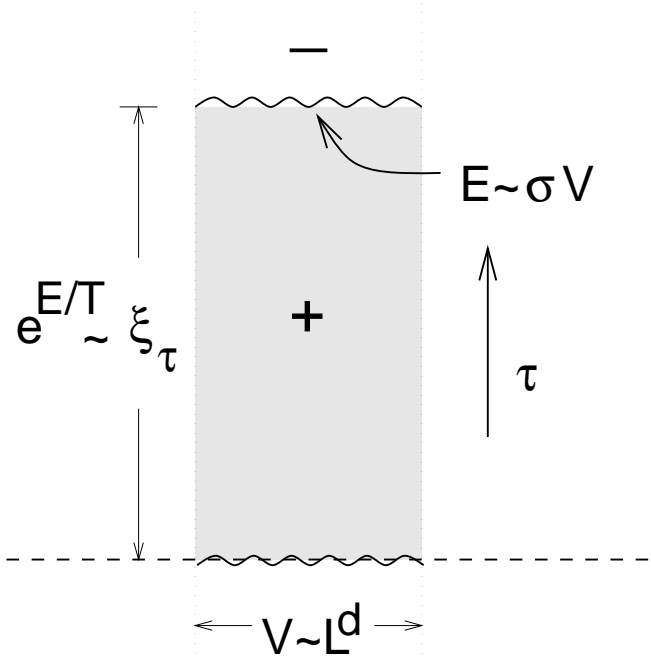}
\caption{\label{sketchfig}
The strongly coupled space region (cluster) of volume $V \sim L^d$ tends to
order (locally) the spins along the imaginary time ($\tau$) direction,
indicated by the plus and minus sign meaning a spin orientation
parallel (plus) or anti-parallel (minus) with respect to the ground
state configuration of the isolated cluster. The insertion of a domain
wall costs an energy $E\sim\sigma V$, where $\sigma$ is a surface
tension (note that the couplings in the $\tau$-direction are all
ferromagnetic). This event occurs with a probability $\exp(-E/T^{\rm cl})$,
resulting in the exponentially large (imaginary) correlation time
$\xi_\tau\sim\exp(\sigma L^d/T^{\rm cl})$.}
\label{fig0}
\end{figure}

In the disordered phase the distributions of
$\chi^{\rm (loc)}$ and $\chi^{\rm (loc)}_{\rm nl}$ will be very broad
with a power law variation at large values. Physically this comes from
regions which are locally ordered.  The probability of having such a
region is exponentially small in its volume, $V$, but, when it occurs,
there is an exponentially large relaxation time\cite{th}, because, to
invert the spins in this region at some imaginary time one has to
insert a domain wall of size $V$, for which the Boltzmann factor is
exponentially small in $V$, as is sketched in Fig.~\protect{\ref{fig0}}.
The combination of an exponentially large
result happening with exponentially small probability gives a
broad distribution of results
in complete analogy to the effect of the Griffiths phase on the
dynamics in {\it classical} random magnets \cite{Randeira,Bray}. In
the latter case the volume to surface ratio determines the resulting
probability distribution for the logarithm of relaxation times. In
contrast to this the extra dimension present in the quantum problem
gives rise to a volume to volume ratio instead (cf.\ Fig.~\ref{fig0}), which
leads to a power law distribution of correlation lengths in the
imaginary time direction. The power depends on the microscopic details
and so is expected to vary smoothly throughout the Griffiths phase.

We can relate the power in the distribution to a dynamical exponent,
defined in the disordered phase, as follows\cite{yr}.  The excitations
which give rise to a large $\chi^{\rm (loc)}$ at $T=0$ are well localized and
so we assume that their probability is proportional to the spatial
volume, $L^d$. These excitations have a very small energy gap, $\Delta
E=E_1-E_0$, where $E_0$ is the ground state energy of the quantum
system and $E_1$ is the first excited state. This small gap
is responsible for the large susceptibility because the
latter is essentially proportional to the inverse of the gap
because the matrix elements
which enter $\chi^{\rm (loc)}$
do not have very
large variations.  Since there is no characteristic energy gap it is
most sensible to use logarithmic variables. Hence, if the power in the
distribution of $\ln \Delta E$ is $\lambda$, say, then we have
\begin{eqnarray}
P( \ln \Delta E) & \equiv & \Delta E \tilde{P}(\Delta E) \nonumber \\
& \sim & L^d \Delta E^\lambda \nonumber \\
& = & \left( L \Delta E^{1/z} \right)^d \ ,
\end{eqnarray}
where the last line defines the dynamical exponent, $z$, in the
conventional way as the power relating a time scale to a length scale.
Comparing the last two expressions we see that

\begin{equation}
\lambda = {d \over z} \ .
\end{equation}

Hence the tail of the distribution of $\Delta E$ has the form
\begin{equation}
\ln \left[ P(\ln \Delta E) \right] = {d\over z} \ln \Delta E + \mbox{const.}
\end{equation}

Since the local susceptibility is proportional to the inverse of the
gap, the power law tail of its distribution should be given by
\begin{equation}
\label{tail_chi}
\ln \left[ P(\ln \chi^{\rm (loc)} ) \right] =
-{d\over z} \ln \chi^{\rm (loc)} + \mbox{const.}
\end{equation}

The non-linear susceptibility involves three integrals over time,
whereas the linear susceptibility only involves one.
Hence we assume that the distribution of
$\chi^{\rm (loc)}_{\rm nl}$ is similar to that of $(\chi^{\rm
(loc)})^3$, which leads to the following power law tail in the
distribution:
\begin{equation}
\label{tail_chi_nl}
\ln \left[ P(\ln \chi^{\rm (loc)}_{\rm nl} ) \right] =
-{d\over 3 z} \ln \chi^{\rm (loc)}_{\rm nl} + \mbox{const.}
\end{equation}
Hence there should be a factor of 3 between the powers in the
distributions of
$\ln \chi^{\rm (loc)}$ and $\ln \chi^{\rm (loc)}_{\rm nl}$. We shall
see that this prediction is confirmed by the numerics. 
As a result, one can characterize {\em all} the Griffiths
singularities in the disordered phase by a {\em single} exponent, $z$.

The {\em average} uniform susceptibility will diverge, in the disordered
phase, at the same point as the average local susceptibility because
spatial correlations are short range and so cannot contribute to a
divergence. From Eq.~(\ref{tail_chi}) we see that this happens when
\begin{equation}
z > d \ .
\label{chi_div}
\end{equation}
Similarly, according to Eq.~(\ref{tail_chi_nl}),
the average non-linear susceptibility will diverge when
\begin{equation}
z > {d \over 3} \ .
\label{chi_div_nl}
\end{equation}

One can also infer the nature of the divergence of $\chi^{\rm
(loc)}_{\rm nl}$ and $\chi^{\rm (loc)}$ as the (real) temperature
$T$ tends to zero, see Fig.~\ref{figm1}. 
For $\chi^{\rm (loc)}$ we expect that the distribution in
Eq.~(\ref{tail_chi}) will be cut off at $\chi^{\rm (loc)} \sim T^{-1}$
which gives
\begin{equation}
[\chi^{\rm (loc)}]_{\rm av} \sim T^{d/ z - 1}.
\end{equation}
For $\chi^{\rm (loc)}_{\rm nl}$ we expect that the cutoff will be
at of order $T^{-3}$, which, together with Eq.~(\ref{tail_chi_nl}) gives
\begin{equation}
[\chi^{\rm (loc)}_{\rm nl}]_{\rm av} \sim T^{d/ z - 3}.
\end{equation}
The global non-linear susceptibility will diverge in the same way,
possibly with logarithmic corrections, as occurs in the
one-dimensional random ferromagnet\cite{dsf}.

The dynamical exponent will tend to some limit as the critical point
is approached.  It is interesting to ask whether this limit will be
the same as the value of $z$ precisely at criticality. On the face of
it, there does not seem any reason why they should be equal, since $z$
in the disordered phase is determined by rare compact clusters,
whereas $z$ at criticality is determined by fluctuations on large
length scales of order of the (divergent) correlation length.
Nonetheless, for the 1-$d$ random ferromagnet, they {\em are} both
equal (to infinity).  For the 2-$d$ spin glass we shall also find that
these two values are numerically close, and may well be equal (though
finite).

\subsection{The Critical Point}
We expect that
the distribution of $\chi^{\rm (loc)}$ will also have a power
law at the critical point just as it does in the disordered phase.
To deduce the exponent, note that
the average time dependent correlation
function at criticality is given by scaling as\cite{ry,gbh}

\begin{equation}
\label{c_tau}
[\langle S_i(0) S_i(\tau) \rangle ]_{\rm av} \sim
{1 \over \tau^{\beta/ (\nu z)} }\ .
\end{equation}
where 
\begin{equation}
{\beta \over \nu} = {d+z-2+\eta \over 2}
\end{equation}
is the order parameter exponent and $\nu$ is the correlation length
exponent. 

Since the average local susceptibility is just the integral of this over
$\tau$, it follows that the
distribution of $\ln \chi^{\rm (loc)}$ must have the same power, i.e.
\begin{equation}
\label{tail_chi_crit}
\ln \left[ P(\ln \chi^{\rm (loc)} ) \right] =
-\left({\beta \over \nu z}\right)
\ln \chi^{\rm (loc)} + \mbox{const.}
\end{equation}
at criticality.
In earlier work\cite{ry} we
found $z=1.5, \eta = 0.5$, so numerically 
$\beta / (\nu z)$ is about 2/3.

Integrating Eq.~(\ref{c_tau}) over $\tau$ from 0 to $T^{-1}$, one sees
that the average susceptibility
(which
is the same as the average local susceptibility for a model with a
symmetric distribution of interactions, such as that used here)
diverges as $T \to 0$ like
\begin{equation}
[\chi]_{\rm av} \sim T^{\beta / z \nu - 1} \ ,
\label{chi_av_crit}
\end{equation}
at criticality. Similarly the average local non-linear susceptibility
will diverge like\cite{ry,gbh}
\begin{equation}
[ \chi^{\rm (loc)}_{\rm nl} ]_{\rm av} \sim
T^{2\beta / z\nu - 3} \ ,
\label{chi_av_nl_crit}
\end{equation}
at criticality.
The global non-linear susceptibility will have a stronger
divergence at criticality\cite{ry}:
\begin{equation}
[ \chi_{\rm nl} ]_{\rm av} \sim
T^{(2\beta - d\nu) / z\nu - 3} \ .
\label{chi_gb_nl_crit}
\end{equation}
Note that Eqs.~(\ref{chi_av_crit}) to (\ref{chi_gb_nl_crit})
refer to the situation in which
$T^{\rm cl}$ is set to the critical value $T^{\rm cl}_c$, and the real
temperature tends to zero, see Fig.~\ref{figm1}. 

\section{results in the disordered phase}

\begin{figure}[hbt]
\epsfxsize=\columnwidth\epsfbox{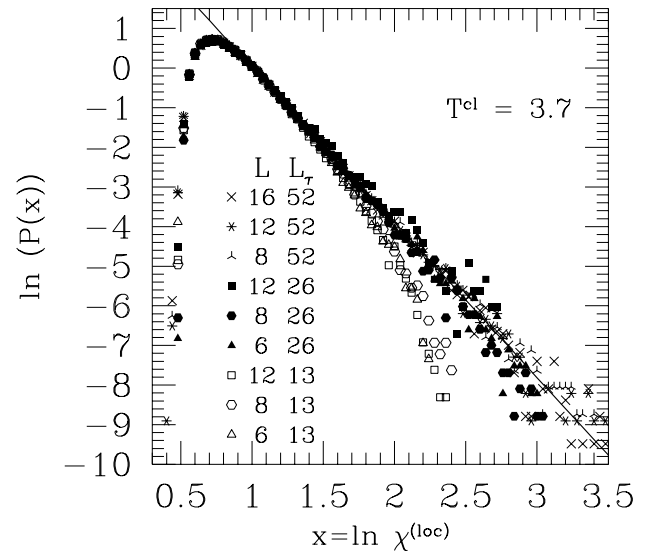}
\caption{
The log of the distribution of the log of the
local susceptibility for $T^{\rm cl} =
3.7$ for different values of $L$ and $L_\tau$.
There is no
significant dependence on $L$ and the data is also independent of
$L_\tau$ at small $\chi^{\rm (loc)}$. Increasing $L_\tau$ seems to simply
extend the range over which the data lies on a straight line. The
solid line
is a fit to the straight line region of the data
and has slope $-3.92$ which gives
$z = 0.51$ from Eq.~(\protect\ref{tail_chi}). 
}
\label{fig1}
\end{figure}

\begin{figure}[hbt]
\epsfxsize=\columnwidth\epsfbox{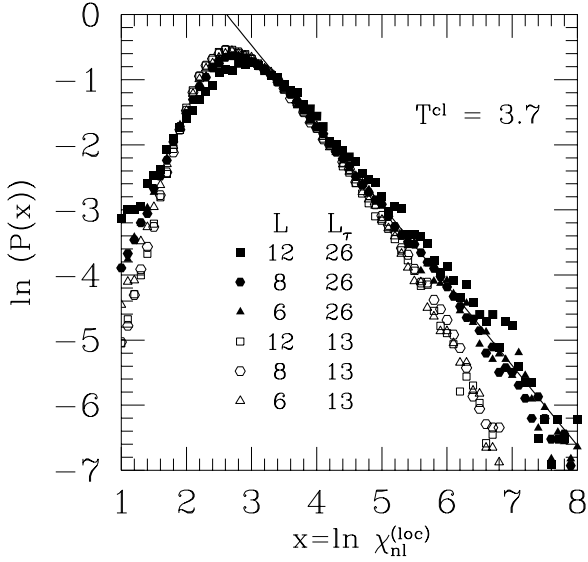}
\caption{
Similar to Fig.~\protect\ref{fig1} but for the local non-linear
susceptibility. The straight line has slope $-1.23$ which gives
$z= 0.54$ from Eq.~(\protect\ref{tail_chi_nl}), in quite good agreement
with the fit to the data in Fig.~\protect\ref{fig1}. Since the slope is
more negative than $-1$, or equivalently $ z < 2/3$, the average non-linear
susceptibility does not diverge at this point. 
}
\label{fig2}
\end{figure}

We use standard Monte Carlo methods to simulate the model in
Eq.~(\ref{ham}). Except where noted, 2560 realizations of the disorder
were averaged over. The simulations were done on parallel computers: a
Parsytec GCel1024 with 1024 nodes (T805 transputers) and a Paragon
XP/S10 with 140 nodes (i860XP microprocessors). Massively parallel machines
with many medium-sized nodes (in terms of memory) are ideal for the
problem considered here: as long as one physical system fits into the
RAM of one processor one only has to set up a farm
topology to distribute the initial seed for the random number
generators and to collect the results
for the different realizations at the end of the simulation. Apart from
that, no communication between processors is needed, so the parallelization
is perfectly efficient; the gain in speed is directly proportional to
the number of nodes.

\begin{figure}[hbt]
\epsfxsize=\columnwidth\epsfbox{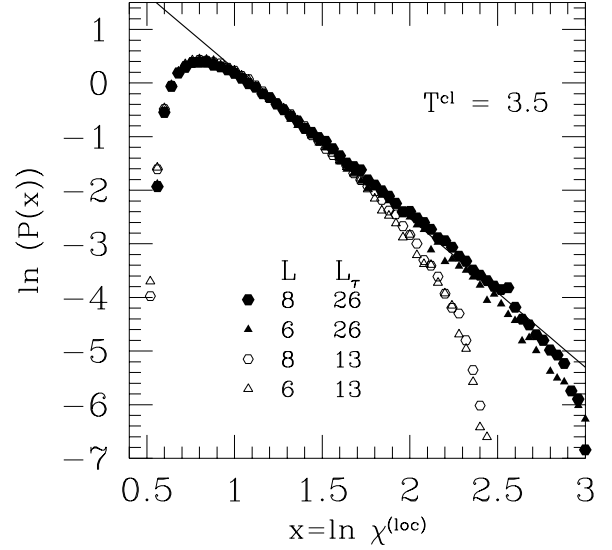}
\caption{
Similar to Fig.~\protect\ref{fig1} but for $T^{\rm cl} = 3.5$.
The straight line has slope $-2.78$ which gives
$z= 0.71$ from Eq.~(\protect\ref{tail_chi}).
}
\label{fig3}
\end{figure}

\begin{figure}[hbt]
\epsfxsize=\columnwidth\epsfbox{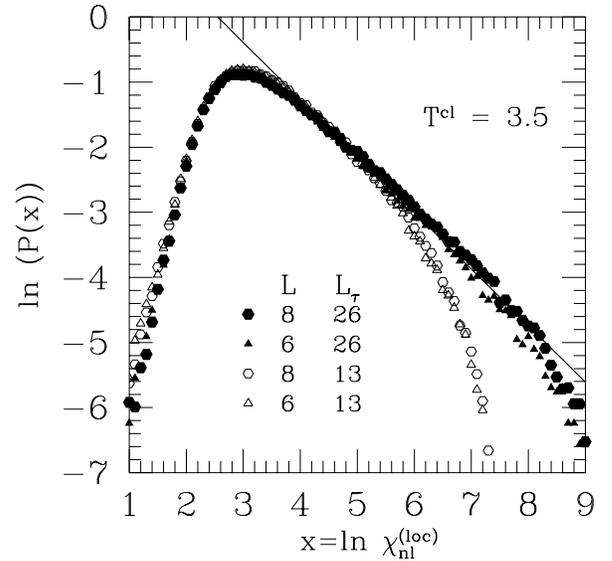}
\caption{
Similar to Fig.~\protect\ref{fig3} but for the local non-linear
susceptibility. The straight line has slope $-0.87$ which gives
$z= 0.76$ from Eq.~(\protect\ref{tail_chi_nl}), in fair agreement
with the fit to the data in Fig.~\protect\ref{fig3}. Since the slope is
greater than $-1$, or equivalently $ z > 2/3$, the average non-linear
susceptibility does diverge at this point. 
}
\label{fig4}
\end{figure}

The distributions of $\ln \chi^{\rm (loc)}$ and $\ln \chi^{\rm
(loc)}_{\rm nl}$ at $T^{\rm cl} = 3.7$ and 3.5 (both in the disordered
phase) are shown in Figs.~\ref{fig1}-\ref{fig4}.  There is a straight
line region for large values as expected, which is independent of $L$,
and the only dependence on $L_\tau$ is that the tail extends further
for larger $L_\tau$. This is not surprising since there is a cutoff
due to the finite number of time slices at $\chi^{\rm (loc)} = L_\tau$ and
$\chi^{\rm (loc)}_{nl} = L_\tau^3$.  At both temperatures one sees
that the values of $z$ obtained from $\chi^{\rm (loc)}$ and $\chi^{\rm
(loc)}_{\rm nl}$ are in reasonably good agreement with each other. At
$T^{\rm cl}= 3.7$, we find that $ z \simeq 0.51$, from the data for
$\chi^{\rm (loc)}$ and $z \simeq 0.54$ from data for $\chi^{\rm
(loc)}_{\rm nl}$. Hence, according to Eq.~(\ref{chi_div_nl}), the
average $\chi^{\rm (loc)}_{\rm nl}$ does not diverge at $T^{\rm cl} =
3.7$ because $z < 2/3$.  At $T^{\rm cl}= 3.5$, we find $z \simeq 0.71$
from the data for $\chi^{\rm (loc)}$ and $z \simeq 0.76$ from data for
$\chi^{\rm (loc)}_{\rm nl}$. Hence, the average $\chi^{\rm
(loc)}_{nl}$ {\em does} diverge at $T^{\rm cl} = 3.5$.

\begin{figure}[hbt]
\epsfxsize=\columnwidth\epsfbox{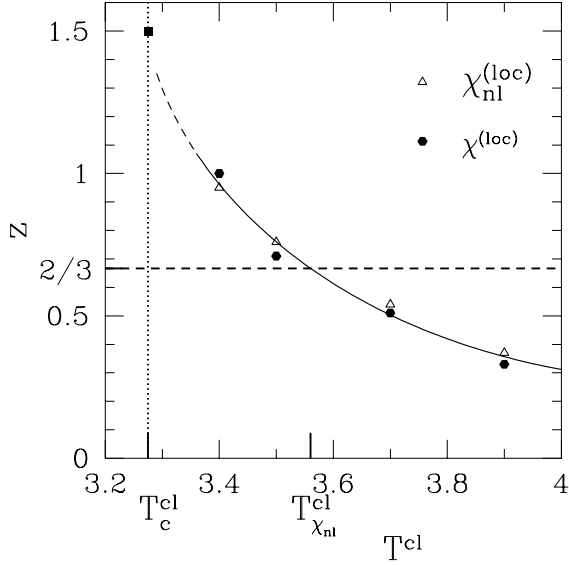}
\caption{
The dynamical exponent $z$, obtained by fitting the distributions of
$\chi^{\rm (loc)}$ and $\chi^{\rm (loc)}_{\rm nl}$ to
Eqs.~(\protect\ref{tail_chi}) and (\protect\ref{tail_chi_nl}),
is plotted for different values of $T^{\rm
cl}$. The estimates obtained from data for
$\chi^{\rm (loc)}$ are shown by the triangles and the estimates from the
data for $\chi^{\rm (loc)}_{\rm nl}$ are shown by the hexagons. The two
are in good agreement. The dotted vertical line indicates the 
critical point, obtained in Ref.~\protect\cite{ry}
and the solid square indicates the estimate of $z$
at the critical point.
The dashed line is $z=2/3$; the
average non-linear susceptibility diverges for $z$ larger than this,
i.e. $T^{\rm cl} > T^{\rm cl}_{\chi_{\rm nl}} \simeq 3.56$. The
solid curve is just a guide to the eye.
}
\label{fig5}
\end{figure}

Fig.~\ref{fig5} shows the values of $z$ at various points in the
disordered phase. In all cases there is good agreement between the
estimates from the data for $\chi^{\rm (loc)}$ and $\chi^{\rm
(loc)}_{\rm nl}$. From this data 
we find that the average non-linear
susceptibility diverges, in the disordered phase, for
\begin{equation}
T^{\rm cl}_c \le T^{\rm cl} \le T^{\rm cl}_{\chi_{\rm nl}} \ ,
\label{tchinldiv}
\end{equation}
where
\begin{equation}
T^{\rm cl}_{\chi_{\rm nl}} \simeq 3.56 \ .
\label{tchi}
\end{equation}
and $T^{\rm cl}_c \simeq 3.275$ from earlier work\cite{ry}.
Note that, according to Fig.~\ref{fig5}
and Eq.~(\ref{chi_div}), the average linear susceptibility does not diverge
anywhere in the disordered phase. 
It appears that the value of $z$ precisely at
criticality may equal the value as the critical point is approached from the
disordered phase, even though it is not clear that they have to be
equal.

\section{results at the critical point}
\begin{figure}[hbt]
\epsfxsize=\columnwidth\epsfbox{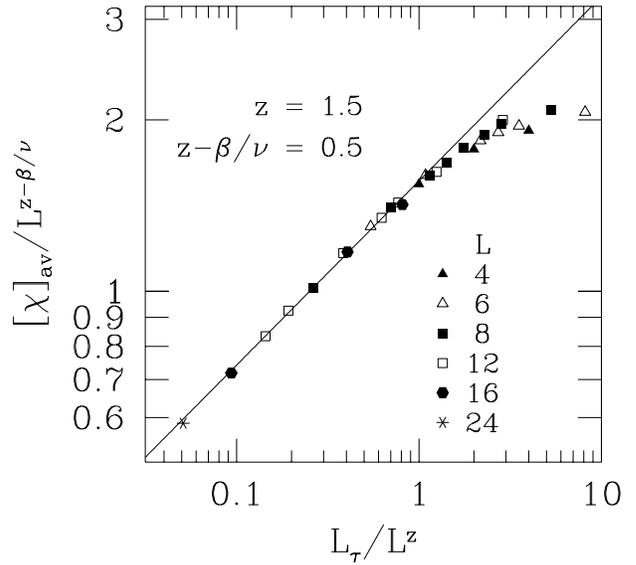}
\caption{
A scaling plot of the data for the average susceptibility at $T^{\rm
cl} = 3.30$, which is the critical point, to within our errors, see
Eq.~(\protect\ref{tc}). The plot assumes the form in
Eq.~(\protect\ref{chi_fss}) with the exponent values deduced in
earlier work\protect\cite{ry}, i.e. $z=1.5, \beta/\nu = 1$. It is seen
that the plot works well. The solid line, which fits the data for
$L_\tau / L^z < 0.8$ has slope $0.5/1.5 = 1/3$ as expected,
since the average susceptibility should be
independent of $L$ in this limit. The power $1/3$ gives the divergence
as $T\to 0$ at criticality, see Eq.~(\protect\ref{t_dep}),
}
\label{fig6}
\end{figure}

From finite size scaling, the average uniform susceptibility (which
is the same as the average local susceptibility for a model with a
symmetric distribution of interactions, such as that used here) varies
with $L$ and $L_\tau$ at the critical point according to\cite{ry,gbh}

\begin{equation}
[\chi ]_{\rm av} = L^x \tilde{\chi} \left({L_\tau \over L^z}\right) \ ,
\label{chi_fss}
\end{equation}
where
\begin{equation}
x = z - {\beta \over \nu} = {z-d+2-\eta \over 2} \ .
\end{equation}
In earlier work\cite{ry} we found $z \simeq 1.5$, $\eta
\simeq 0.5$, $\nu \simeq 1.0$ and $\beta \simeq 1.0$ so
$(z-d+2-\eta)/2 \simeq 1/2$. The earlier work concentrated on an fixed
value of $L_\tau/L^z$, which we call the ``aspect ratio''. Here we
investigate whether Eq.~(\ref{chi_fss}) is satisfied for a range of
aspect ratios. The data, shown in Fig.~\ref{fig6}, does indeed
collapse well with the expected values of the exponents.  For $L_\tau
/ L^z \ll 1$, which corresponds to a large system at finite
temperature, the dependence on $L$ should drop out, and so, from
$L_\tau \propto T^{-1}$, one recovers Eq.~(\ref{chi_av_crit}).
Using the numerical values of the exponents gives
\begin{equation}
[\chi ]_{\rm av}  \sim  T^{-1/3} \ .
\label{t_dep}
\end{equation}
Thus the average
susceptibility diverges at the critical point, though we have seen above
that it does not diverge in the disordered phase. 

\begin{figure}[hbt]
\epsfxsize=\columnwidth\epsfbox{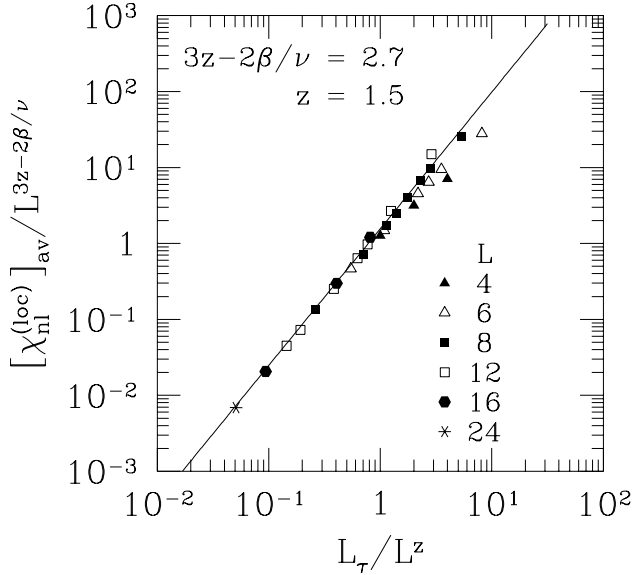}
\caption{
A finite size scaling plot of the average local non-linear susceptibility,
according to Eq.~(\protect\ref{chi_nl_fss}),
at $T^{\rm cl} = 3.30$, very close to the critical point.
The solid line, which fits the data for small
$L_\tau / L^z $ has slope $2.7/1.5 = 1.8$ as expected,
since the average local non-linear
susceptibility should be independent of $L$ in this limit.
The power 1.8 gives the divergence as $T\to 0$, and this value agrees
well with estimates from the earlier estimates of
exponent\protect\cite{ry},
see Eq.~(\protect\ref{t_dep_nl}),
}
\label{fig7}
\end{figure}

The average local non-linear susceptibility is expected to vary at the
critical point as
\begin{equation}
[\chi^{(\rm loc)}_{\rm nl} ]_{\rm av} = L^y \tilde{\chi}^{\rm (loc)}_{\rm nl}
\left({L_\tau \over L^z}\right) \ ,
\label{chi_nl_fss}
\end{equation}
where\cite{ry18}
\begin{equation}
y = 3z - {2\beta \over \nu} \ .
\label{y_def}
\end{equation}

With the numerical values of the exponents found earlier\cite{ry}, one
has $y \simeq 2.5$.  For $L_\tau / L^z \ll 1$, which corresponds to a
large system at finite temperature, the dependence on $L$ should drop
out, and so one recovers Eq.~(\ref{chi_av_nl_crit}).
Using the numerical values of the exponents gives
\begin{equation}
[\chi^{(\rm loc)}_{\rm nl} ]_{\rm av} \sim T^{-5/3} \ .
\label{t_dep_nl}
\end{equation}
The data, shown in Fig~\ref{fig7}, collapses well for $L_\tau/L^z$ not
too large provided $y/z \simeq 1.8$, which gives a $T^{-1.8}$
divergence, close to the prediction in Eq.~(\ref{t_dep_nl}). However,
the data collapse is not as good for larger values of $L_\tau$ with $z
= 1.5$.  Data in this region is difficult to equilibrate, which may be
the cause of the discrepancy. It should be noted, though, that a
better data collapse is obtained for larger values of $z$.  However,
the data for $[\chi]_{\rm av}$ does not scale well with a
significantly larger value of $z$.

\begin{figure}[hbt]
\epsfxsize=\columnwidth\epsfbox{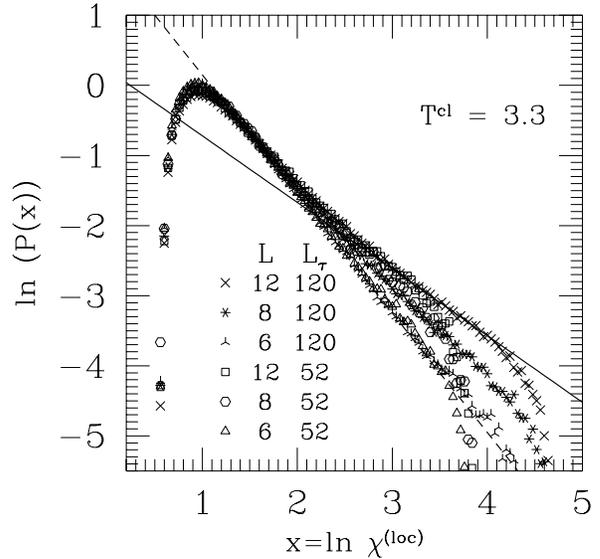}
\caption{
High precision data with large $L_\tau$ for
the distribution of the local susceptibility at the critical point, 
$T^{\rm cl} = 3.30$.
The number of samples was 25600 for $L=6$ and 10240 for $L=8$ and 12.
The dashed line, which is a fit to the
$L=6$ data, has a slope of $-1.7$, and the solid line, which is a fit to the
$L=12$ data, has a slope of $-0.95$. Using the values of exponents found
earlier\protect\cite{ry} the slope, given by
Eq.~(\protect\ref{tail_chi_crit}), is expected to be about $-2/3$ in the
thermodynamic limit, and it is
certainly plausible that data for larger sizes would extrapolate to this
value. 
}
\label{fig8}
\end{figure}

We show results for the distribution of $\chi^{\rm (loc)}$ at the
critical point, $T^{\rm cl} = 3.3$, in Fig.~\ref{fig8}.  Unlike the
data in the disordered phase, shown in Figs.~\protect\ref{fig1} and
\protect\ref{fig3}, there is here a significant size dependence, with
the slope of the tail becoming less negative with increasing
$L$. Asymptotically, the slope should be given by
Eq.~(\ref{tail_chi_crit}), which has the value $-2/3$ using the
exponent values obtained earlier\cite{ry}. The $L=6$ data has a slope
of $-1.7$ and the $L=12$ a slope of $-0.95$ so it is possible that the
slope would tend to $-2/3$ for $L \to \infty$. However, it is also
possible that the slope might be less negative then this, which would
imply a value of $z$ larger than 3/2.

\section{Global non-linear susceptibility}

\begin{figure}[hbt]
\epsfxsize=\columnwidth\epsfbox{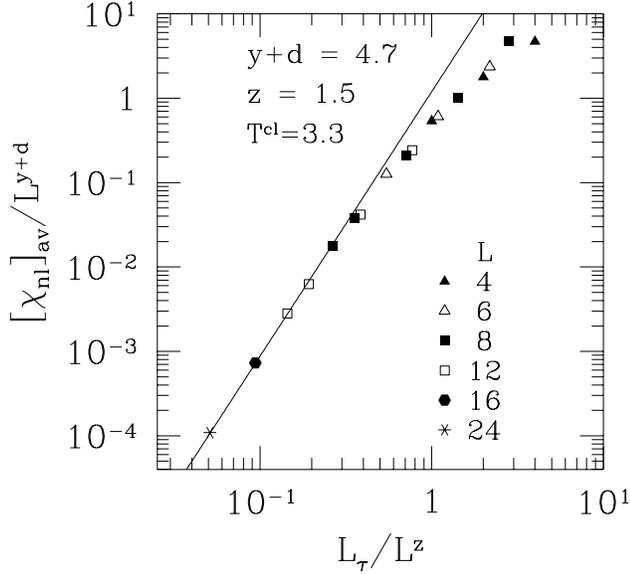}
\caption{
A plot of the average global non-linear susceptibility at criticality,
for a range of sizes and aspect ratios. The exponents used in this
fit are $z=1/5$ and $y+d = 4.7$. 
In earlier work\protect\cite{ry}, we found $z\simeq 1.5$, $y+d \simeq
4.5$ so the present results are consistent with these estimates. In the
limit $L_\tau \ll L^z$, the average global non-linear susceptibility
varies as $L_\tau^{(d+y)/z} \sim T^{-(d+y)/z}$ giving a strong
divergence of roughly $T^{-3.1}$.
This behavior is shown by the solid
line, which is a fit to the data for small $L_\tau/L^z$ and has slope
$3.1$.
}
\label{fig9}
\end{figure}

Experimentally \cite{expt} one measures the {\it global} non-linear
susceptibility, for which the local magnetization $\langle
\sigma_i^z\rangle$ in
(\ref{chinlqm}) and the local external field $h_i$ have to be replaced
by the mean global magnetization,
$L^{-d}\sum_i\langle \sigma_i^z\rangle$
and a uniform field, $H$, respectively.
For the effective classical model this means that one has to consider
\begin{equation}
\chi_{\rm nl}=-\frac{1}{6L_\tau L^d}
\left( \langle M^4 \rangle
- 3 \langle M^2 \rangle^2 \right) 
\end{equation}
with
$M=\sum_{i,\tau}S_i(\tau)$.
In earlier work\cite{ry} we found that
this quantity diverges at criticality (for fixed aspect ratio) like
$\chi_{\rm nl}\sim L^{y+d}$, with $y$ given by (\ref{y_def}), so for
arbitrary aspect ratio, finite size scaling gives
\begin{equation}
[\chi_{\rm nl} ]_{\rm av} = L^{y+d} \tilde{\chi}_{\rm nl}
\left({L_\tau \over L^z}\right) \ ,
\end{equation}
at criticality.
Here, we have looked at scaling of various moments of the global
non-linear susceptibility {\em at the critical point}, for a range of aspect
ratios. For example the average is shown in Fig.~\ref{fig9}. Since there
are two exponents which can be adjusted, $y$ and $z$, the data is unable
to determine them both with precision. However, in 
the limit $L_\tau/L^z \ll 1$, where the dependence on $L$ drops
out and 
\begin{equation}
[\chi_{\rm nl} ]_{\rm av} \sim L_\tau^{(d+y)/z} \sim T^{-(d+y)/z} ,
\end{equation}
the data constrains $(d+y)/z$ to be about 3, in agreement with the $T^{-3}$
divergence at criticality found in earlier work\cite{ry}. Fig.~\ref{fig9}
assumes the previously determined value of $z$, i.e. $z = 1.5$.

\begin{figure}[hbt]
\epsfxsize=\columnwidth\epsfbox{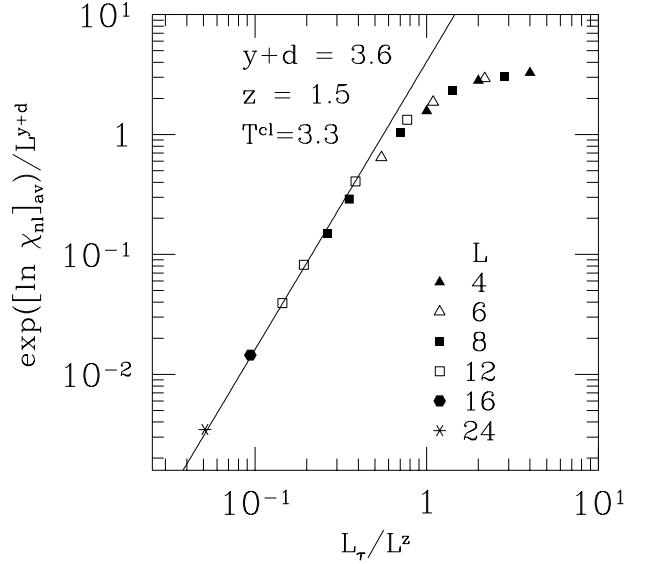}
\caption{
A plot of the typical global non-linear susceptibility,
defined in Eq.~(\protect\ref{chi_typ}), at criticality,
for a range of sizes and aspect ratios. 
In the
limit $L_\tau \ll L^z$, the typical global non-linear susceptibility
varies as $L_\tau^{(d+y)/z} \sim T^{-(d+y)/z}$ giving a quite strong
divergence of roughly $T^{-2.4}$. This behavior is shown by the solid
line, which is a fit to the data for small $L_\tau/L^z$ and has slope
$2.4$.
}
\label{fig10}
\end{figure}

In addition we have evaluated the {\em typical}
global non-linear susceptibility defined by
\begin{equation}
\chi_{\rm nl}^{\rm typ}
=\exp[\log \chi_{\rm nl}]_{\rm av}\;,
\label{chi_typ}
\end{equation}
at the critical point
and show the data in Fig.~\ref{fig10}. As with the data for the average in 
Fig~\ref{fig9}, the ratio $(d+y)/z$ is more tightly constrained than 
either $z$ or $d+y$ separately. A good fit is obtained with
$(d+y)/z \simeq 2.4$, which leads to a divergence of roughly 
$T^{-2.4}$, not quite so strong as from the average non-linear
susceptibility. The difference may well reflect corrections to scaling
for the range of sizes that we were able to study. Note that even the
typical value has a strong divergence with $T$ at criticality, in
contrast to the experiments\cite{expt}. 

We have also studied the probability distribution $P(\chi_{\rm nl})$ in
the disordered phase. This shows a slightly more complicated behavior
than the probability distribution of local quantities presented above.
We find that the power describing the tail in the same distribution is
the same as that of the local non-linear susceptibility. This is
reasonable since these unusually large values come from correlations
which are very long ranged in time, whereas the spatial correlation
length is small and so does not give a significant extra effect. These
spatial correlations do, however, cause the peak in the distribution to
shift to larger values with increasing system size, though presumably
the peak position would eventually settle down to a constant for sizes
greater than the correlation length.

\section{conclusions}
One can characterize Griffiths singularities in the disordered phase of
a quantum system undergoing a $T=0$ transition with discrete broken
symmetry, by a single, continuously varying, dynamical exponent, $z$.
Average response functions may or may not diverge in part of the
disordered phase near the critical point, depending on the value of
$z$, see Eqs.~(\ref{chi_div}) and (\ref{chi_div_nl}).  The numerical
results, summarized in Fig.~\ref{fig5} and Eqs.~(\ref{tchinldiv}) and
(\ref{tchi}), indicate that the average linear susceptibility does not
diverge in the disordered phase of the 2-$d$ quantum Ising spin glass,
though it diverges at the critical point. The average non-linear
susceptibility {\em does}, however, diverge in part of the disordered
phase.

Numerically, as one approaches the critical point from the disordered
phase, the value of $z$ is close to the value obtained precisely at the
critical point, see Fig.~\ref{fig5}. Since the same result is known to
hold exactly in 1-$d$, where they are both equal to infinity\cite{dsf},
one might 
speculate that they are equal in general, though we are not aware of any
proof of this. 
Presumably the detailed dependence of $z$ with $T^{cl}$ in the
disordered phase, shown in Fig.~\ref{fig5}, is non-universal. However,
it is interesting to ask whether the answer to the question ``{\em Does
the non-linear susceptibility diverge in the disordered phase}'' is
universal or not.  From Eq.~(\ref{chi_div_nl}) this depends on whether
$z > d/3$ as the critical point is approached. If this limit for $z$ is
precisely the same as the value of $z$ at criticality,
then the answer to the question {\em is}
universal.  However, as we just mentioned,
we are not aware of any argument which shows that
these two values of $z$ should be equal in general.

A related study has also been carried out recently by Guo et
al.\cite{gbh2,g_thesis} Their results for the 2-$d$ spin glass are consistent
with ours, and they also performed some simulations for the 3-$d$ spin
glass. In three dimensions the classical model has a finite temperature
transition\cite{review}, so the spin glass phase would exist for a
finite range of $T$, which shrinks to zero as $T^{\rm cl} \to {T^{\rm
cl}_c}^-$, see Fig.~\ref{figm1}. The results of Guo et al.\cite{gbh2}
indicate that, in three dimensions,
the range of the Griffiths phase over which the
non-linear susceptibility diverges, i.e. the region between $T^{\rm
cl}_c$ and $T^{\rm cl}_{\chi{\rm nl}}$ in Fig.~\ref{figm1}, is very
small but apparently non-zero.
It is interesting to speculate on whether the possible
divergence of the non-linear susceptibility in part of the disordered
phase might be related to the difference between the
experiments\cite{expt} which apparently do not find a strong divergence
of $\chi_{\rm nl}$ at the quantum critical point, and the
simulations\cite{ry,gbh} which do.

The data presented here is consistent with our earlier results\cite{ry}
in finding a dynamic exponent at the critical point of about 1.5. This
is rather different from the situation in one-dimension\cite{sm,dsf}
where $z = \infty$, and one might ask whether the true dynamical
exponent might not be larger than 1.5 in two-dimensions, and possibly
infinite. While the data for the modest range of sizes that can be
studied by Monte Carlo simulations is consistent with a small value of
$z$, we cannot completely rule out the possibility that this estimate
would increase if one could study larger sizes. Unfortunately, it does
not seem feasible to study very much larger sizes with current computer
power, unless a more sophisticated algorithm can be found than the
single spin-flip approach used here. Assuming that $z$ is indeed
finite, the critical scaling in the two-dimensional quantum Ising spin
glass is of a fairly conventional, but anisotropic, type, with $z$
playing the role of an anisotropy exponent. The difference from a
classical magnet with anisotropic scaling is that Griffiths
singularities give additional singularities in various scaling
functions in the limit $L_\tau \gg L^z$, or equivalently $T L^z \ll
1$.

\acknowledgments
The work of APY has been supported by the National Science Foundation
under grant No. DMR--9411964. The work of HR was supported by the
Deutsche Forschungsgemeinschaft (DFG) and he thanks the Physics
Department of UCSC for the kind hospitality. We should like to thank
R.~N.~Bhatt, M.~Guo, D.~A.~Huse and D.~S.~Fisher for helpful
discussions.

\end{document}